\documentstyle[epsfig]{aipproc}
\begin{document}
\title{Unsupervised Induction and Gamma-Ray Burst Classification }

\author{Richard J. Roiger$^{\dagger}$ 
  Jon Hakkila$^{\dagger}$, David J. Haglin$^{\dagger}$ \\
  Geoffrey N. Pendleton$^{\ddag}$, Robert S. Mallozzi$^{\ddag}$
}

\address{%
   $^{\dagger}$Minnesota State University, Mankato, MN, 56001 \\
   $^{\ddag}$University of Alabama in, Huntsville, AL, 35812
}

\maketitle

\begin{abstract}
We use ESX, a product of Information Acumen Corporation, to 
per\-form
unsupervised learning on a data set containing 797 gamma-ray 
bursts taken from the BATSE 3B catalog\cite{m96}. Assuming all 
attributes to be distributed logNormally, Mukherjee et al.\cite{mfbmfr98}
analyzed these same data using a statistical cluster analysis. Utilizing
the logarithmic values for T90 duration, total fluence, and hardness ratio
HR321 their results showed the instances formed three classes. Class I
contained long/bright/intermediate bursts, class II consisted of
short/faint/hard bursts and class III was represented by 
intermediate/intermediate/soft bursts. 

When ESX was presented with these data and restricted to forming a small 
number of classes, the two classes found by previous standard techniques
\cite{cd} were determined. However, when ESX was allowed to form more
than two classes, four classes were created. One of the four classes contained a 
majority of short bursts, a second class consisted of mostly intermediate 
bursts, and the final two classes were subsets of the Class I (long) bursts 
determined by Mukherjee et al.  We hypothesize that systematic biases may
be responsible for this variation. 
\end{abstract}

\section*{Introduction}

Induction-based learning\cite{ls95} attempts to extract interesting 
patterns from data. These patterns form concept classes with each class
containing data instances. When the induction is unsupervised, the
learning model has no a priori class knowledge. Rather, the learning
algorithm uses one or more statistical or symbolic (machine learning)
evaluation functions to cluster instances into concept classes. 

Mukherjee et al.\cite{mfbmfr98} performed a statistical cluster analysis
on a data set containing 797 gamma-ray bursts taken from the BATSE 3B
catalog\cite{m96}. Assuming all attributes to be distributed logNormally, and
utilizing the logarithmic values for T90 duration, total fluence, and
hardness ratio HR321 their results showed the instances formed three
classes. Class I contained long/bright/intermediate bursts, class II
consisted of short/faint/hard bursts and class III was represented by
intermediate/intermediate/soft bursts. Table \ref{table1} shows the mean
and standard deviation values for the three classes.
Table \ref{table2} offers a best defining rule for each class, as
determined by ESX\cite{rghh99}. The rule for class I bursts
indicates that 82.72\% of the bursts in this class have a log T90 value between
.70  and 2.66 and a log Fluence between -5.77 and -3.11.  The rule also shows
that we can be at least 97\% confident that a burst with these characteristics
is a class I burst. Table \ref{table2} shows that the class III rule does
not cover its instances as well as the rules for classes I and II.  

\newlength{\columnWidth}
\columnWidth = 6mm

\begin{table}[ht!]
\caption{Mean and Standard Deviations for the Classes found by
     Mukherjee et al. (1998)}
\begin{tabular}{|r|*{4}{@{\hspace{\columnWidth}}c@{\hspace{\columnWidth}}|}}
\noalign{\vspace{-8pt}}
 & \underline{Class I} & \underline{Class II} & \underline{Class III} &
      \bf{Domain} \\
 & Long & Short & Intermediate & \\ \hline 
Number of Bursts & 486 & 203 & 107 & 796 \\ \hline
\tablenote{Attributes log T50, log P256, and log HR32 were not used in the
  final analysis since each had a high correlation with its respective
  counterpart (log T90, log fluence, and log H321).}
Log T50 (mean) & 1.13 & -0.80 & 0.33 & 0.53 \\ \hline
(sd) & 0.45 & 0.41 & 0.26 & 0.93 \\ \hline
Log T90 (mean) & 1.55 & -0.42 & 0.71 & 0.93 \\ \hline
(sd) & 0.40 & 0.44 & 0.32 & 0.94 \\ \hline
Log Fluence (mean) & -5.21 & -6.37 & -6.11 & -5.63 \\ \hline
(sd) & 0.59 & 0.57 & 0.37 & 0.77 \\ \hline
$^{\rm a}$ Log P256 (mean) & 0.21 & 0.14 & -0.08 & 0.15 \\ \hline
(sd) & 0.48 & 0.38 & 0.33 & 0.45 \\ \hline
$^{\rm a}$ Log HR32 (mean) & 0.20 & 0.51 &  0.09 & 0.26 \\ \hline
(sd) & 0.27 & 0.27 & 0.40 & 0.33 \\ \hline
Log HR321 (mean) & 0.43 & 0.70 & 0.35 & 0.49 \\ \hline
(sd) & 0.23 & 0.26 & 0.39 & 0.30 \\
\end{tabular}
\label{table1}
\end{table}

\begin{table}[ht!]
\caption{Representative ESX Rules for the Three Classes found by Mukerjee
et al. (1998)}
\begin{tabular}{|c|l|} 
\noalign{\vspace{-8pt}}
              & $0.70 <= \log {\rm T90} <= 2.66$ \\
Class I       & and $-5.77 <= \log {\rm Fluence} <= -3.11$ \\
(Long Bursts) & \quad :rule accuracy 97.34\% \\
              & \quad :rule coverage 82.72\% \\ \hline
   Class II    & $-1.55 <= \log \rm{T90} <= 0.41$ \\
(Short Bursts) & \quad :rule accuracy 90.87\% \\
               & \quad :rule coverage 98.03\% \\ \hline
Class III      & $0.46 <= \log \rm{T90} <= 0.96$ \\
(Intermediate  & and $0.17 <= \log \rm{T50} <= 0.55$ \\
Bursts)        & \quad :rule accuracy 79.17\% \\
               & \quad :rule coverage 53.27\% \\ \hline
\end{tabular}
\label{table2}
\end{table}

In this paper we use ESX\cite{rghh99}, a machine learning model  and product of 
Information Acumen Corporation, to perform unsupervised learning on these same
data for the purpose of comparative analysis. We chose ESX for this research
since ESX explains its behavior has been shown to perform well in several
real-world environments \cite{rghh99}.   

\section*{Method}

The machine learning component of ESX is an induction-based sequential learning
model that creates a concept hierarchy\cite{glf89} from a set of input
instances.  ESX uses knowledge contained in its concept hierarchy to
generate a set of production rules to help define and explain what has
been discovered. Supervised as well as unsupervised learning is supported.

ESX accepts data in the form of instances represented in attribute-value format.
When learning is unsupervised, ESX takes one of two possible actions for each
newly presented instance: (1) Place the new instance into an existing
cluster, or (2) create a new conceptual cluster containing the instance as
its only member.  
 
In addition, ESX allows the user to set a learning parameter so as to
encourage or discourage the creation of new clusters. For a given domain, a
best value for this parameter can be determined experimentally.
 
\section*{Results}

For our first experiment, we set the ESX learning parameter so as to restrict
the formation of new classes.  As a result, ESX clustered the data into the
two classes found by previous standard techniques \cite{cd}. Table
\ref{table3} shows a representative rule for each class. Notice that both
clusters are well-defined. 

\begin{table}[ht!]
\caption{Representative Rules Taken from the Two Class ESX Clustering} 
\begin{tabular}{|c|l|}
\noalign{\vspace{-8pt}}
   Class I    & $0.54 <= \log {\rm T90} <= 2.66$ \\   
(Long Bursts) & \quad :rule accuracy 98.03\% \\
              & \quad :rule coverage 96.99\% \\ \hline
   Class II    & $-1.55 <= \log \rm{T90} <= 0.38$ \\
(Short Bursts) & \quad :rule accuracy 98.14\% \\
               & \quad :rule coverage 90.95\% \\ \hline
\end{tabular}
\label{table3}   
\end{table}

For our second experiment, we allowed ESX to form a best set of three or
more clusters. The results of this experiment showed the formation of four
clusters. One of the four clusters contained a majority of intermediate
bursts (class 1); a second cluster consisted of mostly short bursts (class
2). The remaining two clusters (classes 3 and 4) were subsets of the
Mukherjee class I bursts.  The class mean and standard deviation values
for each of the six burst attributes are shown in Table \ref{table4}. 

\columnWidth = 3mm

\begin{table}[ht!]
\caption{Mean and Standard Deviations for the ESX Four Class Clustering}
\begin{tabular}{|r|*{5}{@{\hspace{\columnWidth}}c@{\hspace{\columnWidth}}|}}
\noalign{\vspace{-8pt}}
 & \underline{Class 1} & \underline{Class 2} & \underline{Class 3} &
      \underline{Class 4} & \bf{Domain} \\
 &Intermediate & Short & Long/Soft & Long/Bright & \\ \hline 
Number of Bursts & 182 & 205 & 195 & 215 & 796 \\ \hline
Log T50 (mean) & 0.44 & -0.78 & 1.27 & 1.18 & 0.53 \\ \hline
(sd) & 0.44 & 0.44 & 0.37 & 0.44 & 0.93 \\ \hline
Log T90 (mean) & 0.85 & -0.41 & 1.67 & 1.62 & 0.93 \\ \hline
(sd) & 0.37 & 0.46 & 0.32 & 0.38 & 0.94 \\ \hline
Log Fluence (mean) & -5.87 & -6.36 & -5.50 & -4.84 & -5.63 \\ \hline
(sd) & 0.45 & 0.59 & 0.37 & 0.61 & 0.77 \\ \hline
Log P256 (mean) & 0.04 & 0.13 & -0.07 & 0.48 & 0.15 \\ \hline
(sd) & 0.43 & 0.38 & 0.22 & 0.51 & 0.45 \\ \hline
Log HR32 (mean) & 0.11 & 0.54 &  -0.03 & 0.38 & 0.26 \\ \hline
(sd) & 0.27 & 0.30 & 0.24 & 0.16 & 0.33 \\ \hline
Log HR321 (mean) & 0.36 & 0.73 &  0.24 & 0.59 & 0.49 \\ \hline
(sd) & 0.27 & 0.29 & 0.21 & 0.14 & 0.30 \\
\end{tabular}
\label{table4}
\end{table}
Table \ref{table5} offers representative rules for each of the four
clusters.  Figures 1 and 2  as well as Table \ref{table4} indicate that
class 3 contains mostly long/soft bursts and class 4 contains long/bright
bursts. The following rule represents a covering rule for the cluster
formed by combining the class 3 and class 4 bursts. 

\begin{center}
\begin{tabular}[ht!]{l}
$1.19 <= \log \rm{T90} <= 2.66$ \\
\quad :rule accuracy 90.26\% \\
\quad :rule coverage 92.68\%
\end{tabular}
\end{center}

\begin{table}[ht!]
\caption{Representative Rules Taken from the Four Class ESX Clustering}
\begin{tabular}{|c|p{2.0in}|p{2.0in}|}
\noalign{\vspace{-8pt}}
	       & $0.29 <= \log \rm{T90} <= 1.20$ & 
                 $0.21 <= \log \rm{T50} <= 0.63$ \\
Class 1        & \quad :rule accuracy 75.00\% &
                 and $0.29 <= \log \rm{T90} <= 1.09$ \\
(Intermediate) & \quad :rule coverage 74.18\% &
                 \quad :rule accuracy 89.02\% \\
               & & \quad :rule coverage 40.11\% \\ \hline
               & $-1.55 <= \log \rm{T90} <= 0.42$ & 
                 $-7.80 <= \log \rm{Fluence} <= -6.63$ \\
Class 2        & and $-1.92 <= \log \rm{T50} <= -0.02$ &
                 and $-1.92 <= \log \rm{T50} <= -0.02$ \\
(Short)        & \quad :rule accuracy 93.20\% &
                 \quad :rule accuracy 95.95\% \\
               & \quad :rule coverage 93.66\% &
                 \quad :rule coverage 34.63\% \\ \hline              
Class 3        & & $0.02 <= \log \rm{HR321} <= 0.08$ \\
(Long/Soft)    & $1.19 <= \log \rm{T90} <= 2.66$ &
                 \quad :rule accuracy 77.36\% \\
               & \quad :rule accuracy 90.26\% &
                 \quad :rule coverage 21.03\% \\ \cline{1-1} \cline{3-3}
Class 4        & \quad :rule coverage 92.68\% &
                 $-4.85 <= \log \rm{Fluence} <= -3.11$ \\
(Long/Bright)  & & \quad :rule accuracy 90.91\% \\
               & & \quad :rule coverage 51.16\% \\
\end{tabular}
\label{table5}
\end{table}

\section*{Conclusions}

We used ESX to cluster data about 797 gamma ray bursts.  When 
restricted to forming a small number of classes,
two classes were determined.
However, when allowed to form more
than two classes, four classes were created.
Two of the clusters were similar
to the class II and class III bursts 
determined by Mukherjee et al.\cite{mfbmfr98}. 
Taken together, the two
remaining clusters represent the class I Mukherjee et al. bursts.
ESX differentiated the class I bursts by 
brightness and hardness.
The separation of long
bursts into two classes may be due in part to the fact that 
ESX makes no a priori assumptions about data distribution.    

We hypothesize that systematic effects may cause some class I bursts to
take on class III characteristics\cite{Hak}.
Systematic
biases may explain why class I bursts have been separated into two groups
by ESX.  Our future work will focus on testing these hypotheses with the
help of additional 
induction-based techniques.

\newpage

\begin{figure}[hb]
\centerline{\epsfig{file=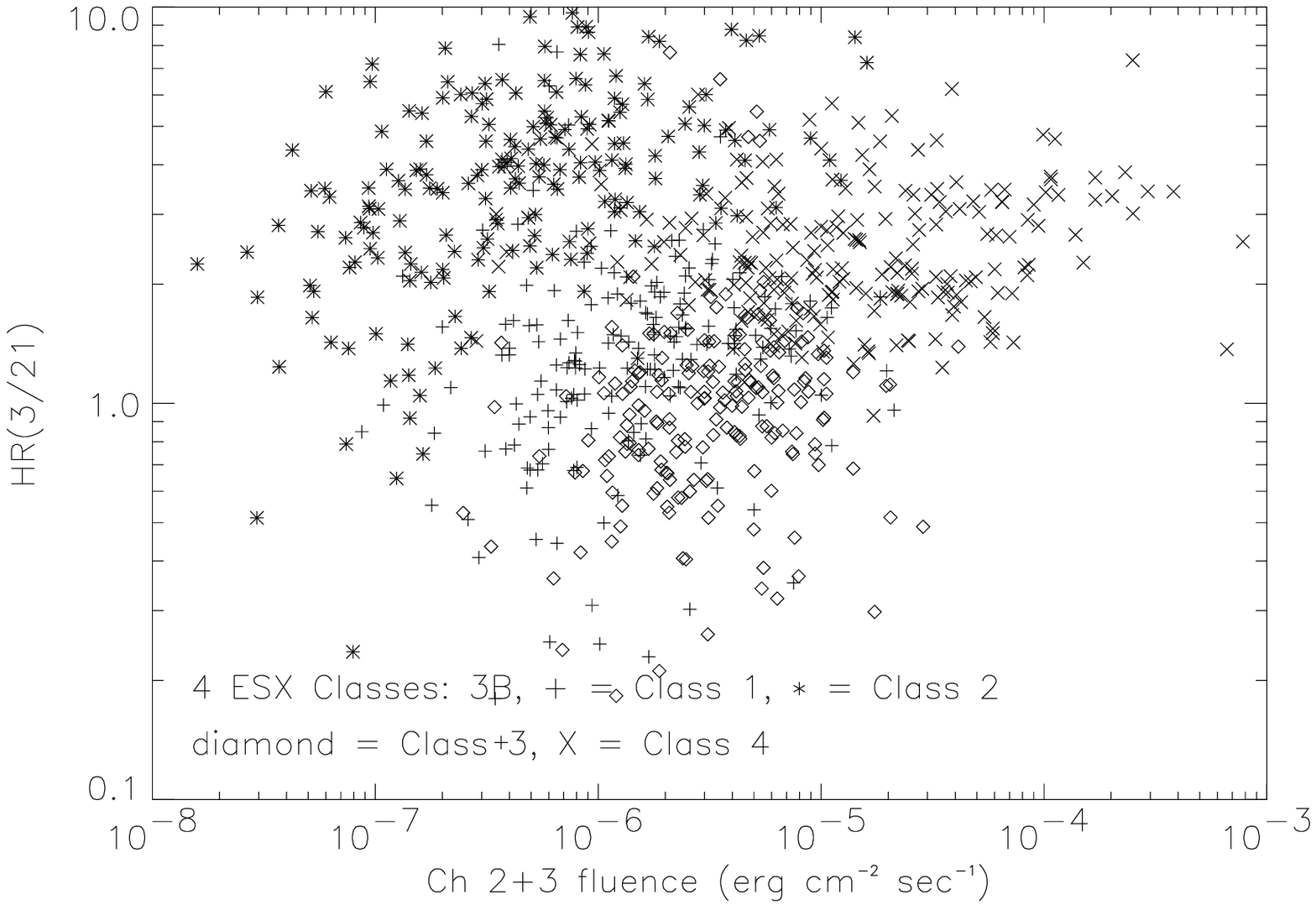,width=3.5 in}}
\vspace{10pt}
\caption{3/21 Hardness Ratio vs. ch 2 + 3 fluence}
\label{fig:ST90}

\centerline{\epsfig{file=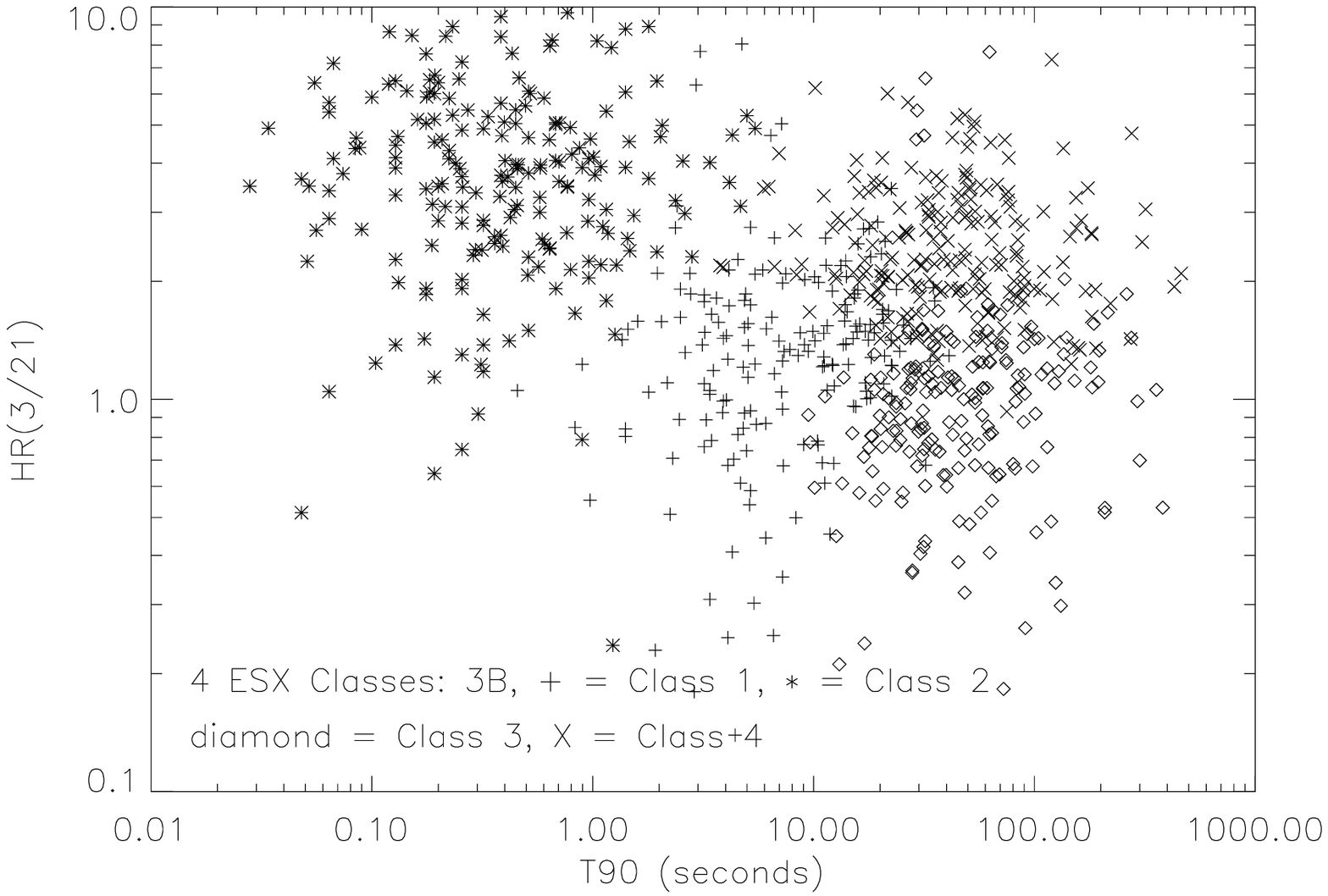,width=3.5 in}}
\vspace{10pt}
\caption{3/21 Hardness Ratio vs. T90 duration}
\label{fig:hr321S}
\end{figure}

\end{document}